\documentclass[aps,prl,twocolumn,superscriptaddress,floatfix,notitlepage,10pt]{revtex4-2}

\usepackage{graphicx}
\usepackage[breaklinks,colorlinks,urlcolor=blue,citecolor=blue,linkcolor=magenta]{hyperref}
\usepackage{amsmath,amsfonts,amssymb}
\usepackage{xcolor}
\usepackage{orcidlink}
\usepackage{comment}
\usepackage[normalem]{ulem}

\newcommand{\fb}{\bar{\phi}}
\newcommand{\VFSB}{V_\text{B}}

\newcommand{\Mpl}{M_\text{Pl}}
\newcommand{\ft}{\tilde{\phi}}
\newcommand{\Tosc}{T_\star}
\newcommand{\Trh}{T_\text{rh}}
\newcommand{\Teq}{T_\text{eq}}

\definecolor{darkblue}{rgb}{0,0.1,0.5}
\definecolor{darkgreen}{rgb}{0,0.5,0.2}
\definecolor{darkred}{RGB}{153,26,0}
\definecolor{seablue}{rgb}{0,0.2,0.6}
\definecolor{light}{rgb}{0,0.2,0}
\definecolor{viola}{RGB}{134,41,198}
\definecolor{myGreen}{RGB}{20,140,0}

\begin{document}

\title{Ultralight Dark Matter from the Edge of Field Space}

\author{Mathias Becker}
\email{mathias.becker@unipd.it}
\affiliation{Dipartimento di Fisica e Astronomia, Universit\`{a} degli Studi di Padova, Via Marzolo 8, 35131 Padova, Italy}
\affiliation{INFN, Sezione di Padova, Via Marzolo 8, 35131 Padova, Italy}

\author{Francesco D'Eramo}
\email{francesco.deramo@pd.infn.it}
\affiliation{Dipartimento di Fisica e Astronomia, Universit\`{a} degli Studi di Padova, Via Marzolo 8, 35131 Padova, Italy}
\affiliation{INFN, Sezione di Padova, Via Marzolo 8, 35131 Padova, Italy}

\author{Ville Vaskonen}
\email{ville.vaskonen@pd.infn.it}
\affiliation{Dipartimento di Fisica e Astronomia, Universit\`{a} degli Studi di Padova, Via Marzolo 8, 35131 Padova, Italy}
\affiliation{INFN, Sezione di Padova, Via Marzolo 8, 35131 Padova, Italy}
\affiliation{Keemilise ja bioloogilise f\"u\"usika instituut, R\"avala puiestee 10, 10143 Tallinn, Estonia}

\begin{abstract}
We introduce a novel class of bosonic dark matter candidates that we dub wallions, featuring boundaries in field space. The wallion mass is exponentially suppressed when the separation between boundaries far exceeds their intrinsic width and remains radiatively stable under self-interactions. We study the early-universe evolution of wallions and the associated cosmological signatures. Finally, we show that instanton effects can dynamically generate field-space boundaries and discuss possible experimental probes once the wallion couples to Standard Model fields.
\end{abstract}

\maketitle

\noindent {\bf Introduction.} The observational evidence for dark matter (DM)~\cite{Jungman:1995df,Bertone:2004pz,Feng:2010gw,Planck:2018vyg,Arbey:2021gdg,Cirelli:2024ssz,Bozorgnia:2024pwk} leaves no doubt about the need for physics beyond the Standard Model (SM). Decades after its first indications, the breadth of empirical evidence is matched only by the diversity of theoretical proposals aiming to explain its microscopic particle nature.

Searches for DM have predominantly targeted weakly interacting massive particles (WIMPs), motivated by their connection to the hierarchy problem~\cite{Gildener:1976ai,Weinberg:1978ym,Giudice:2008bi} and by the fact that their relic abundance naturally emerges from thermal freeze-out~\cite{Lee:1977ua,Goldberg:1983nd,Scherrer:1985zt,Srednicki:1988ce,Gondolo:1990dk}. The persistent absence of WIMP signals in recent years has placed this framework under pressure~\cite{Arcadi:2017kky,Roszkowski:2017nbc,Arcadi:2024ukq}. Meanwhile, an extensive experimental effort~\cite{Ehret:2010mh,CAST:2008ixs,CAST:2017oph,IAXO:2019mpb,ADMX:2009iij,ADMX:2020xuj,HAYSTAC:2021rbt,Jeong:2020cwz,ABRACADABRA:2020gpx,DMRadio:2022dft,SHAFT:2023xhb,ORGAN:2023zms}, together with theoretical advances~\cite{Hu:2000ke,Arvanitaki:2009fg,Marsh:2010wq,Graham:2011qk,Svrcek:2006yi}, has reignited interest in ultralight and extremely weakly coupled bosonic candidates. The QCD axion~\cite{Wilczek:1977pj,Weinberg:1977ma,Preskill:1982cy,Abbott:1982af,Dine:1982ah}, introduced to solve the strong CP problem via the Peccei–Quinn mechanism~\cite{Peccei:1977hh,Peccei:1977ur}, is a particularly motivated example. WIMPs and axions, two leading DM paradigms with deep connections to open questions in particle physics, delineate two distinct mass regimes: \emph{particle} DM, with masses above the eV scale, and \emph{wave} DM, associated with lighter candidates whose de~Broglie wavelengths are macroscopic.

Ultralight wave DM has emerged as a particularly compelling candidate~\cite{Arias:2012az,Marsh:2015xka,Ferreira:2020fam}. It leaves distinctive imprints in cosmological observables, notably the suppression of small-scale structure~\cite{Hu:2000ke,Schive:2014dra,Mocz:2017wlg}, and has motivated a new generation of precision laboratory experiments probing its subtle effects~\cite{VanTilburg:2015oza,Graham:2015ifn,Hees:2016gop,Derevianko:2016vpm,Safronova:2017xyt,Irastorza:2018dyq}. On astrophysical scales, ultralight DM can induce striking phenomena such as black hole superradiance~\cite{Arvanitaki:2010sy,Brito:2015oca,Baryakhtar:2017ngi,Baumann:2018vus,Ng:2020ruv}, offering additional avenues to constrain its properties. Theoretical consistency, however, requires an explanation for the origin of such tiny masses and their stability against quantum corrections. The QCD axion provides an instructive example: its mass arises dynamically after confinement, while its smallness is protected by its Nambu--Goldstone nature.

In this Letter, we introduce a novel candidate in the landscape of ultralight DM, the \emph{wallion}, whose mass is naturally protected by limits of its field space. Unlike axions, which live on a compact manifold with periodic boundary conditions, the wallion~$\phi$ roams freely up to a hard boundary. Such frameworks have recently attracted attention in high-energy physics~\cite{Cheung:2024wme,Nicolis:2022llw,Borghetto:2025jrk}, where field space itself becomes dynamical.
Ref.~\cite{Cheung:2024wme} studied scalar effective field theories (EFTs) in which the potential rises sharply as the field approaches a limiting value $|\phi|\simeq \fb$. The parameter $\fb$ therefore marks a \emph{field-space boundary}: field excursions with $|\phi|\gtrsim \fb$ become energetically costly, effectively confining dynamics to the interior region $|\phi|<\fb$. Remarkably, it was shown that this structure is preserved under quantum corrections. In the second half of this Letter, we discuss how scalar potentials of this kind can arise from instanton dynamics.

What makes these scenarios interesting in the context of wave DM? Let us first consider the most extreme, albeit unrealistic, case of a potential with a wall:  
\begin{equation} \label{eq:Vwall}
    V_\text{wall}(\phi) = 
    \begin{cases}
        0, & \quad |\phi| < \fb \\
        \infty, & \quad |\phi| \geq \fb
    \end{cases} \, .
\end{equation}
Within the well the potential is flat and the field massless, $m_\phi^2=0$.  Since realistic potentials are continuous, the wall potential must be regularized. Ref.~\cite{Cheung:2024wme} provides an example with a potential of the form
\begin{equation} \label{eq:VFSB}
    \VFSB (\phi) = \Lambda^4 \exp\left[ \frac{\phi^2 - \fb^2}{\Lambda^2}\right] + V_0 \, .
\end{equation}
The parameter $\Lambda$, which has mass dimension one, can be interpreted as the wall thickness. The field-space boundary $\fb$ is identified with the field value at which the scalar potential, excluding the constant contribution $V_0$, reaches the EFT scale, $V_B(\fb)-V_0=\Lambda^4$~\footnote{Field-space boundaries arise more generally within EFTs of the form 
$V(\phi)=\epsilon\,\Lambda^4\sum_{n=0}^\infty (c_n/n!)(\phi^2/\Lambda^2)^n$, with $\epsilon\ll1$~\cite{Cheung:2024wme}. 
The potential in Eq.~\eqref{eq:VFSB}, up to the constant $V_0$, corresponds to $c_n=1$ and $\epsilon=\exp(-\fb^2/\Lambda^2)$. 
The field-space boundary is defined as the field value $\phi=\fb$ for which the dominant term in the series becomes $\mathcal{O}(\Lambda^4)$, signaling the breakdown of the small-$\epsilon$ suppression. 
For coefficients $c_n=\mathcal{O}(1)$ this yields $\fb\simeq\Lambda\sqrt{-\ln\epsilon}$, while for the exponential potential in Eq.~\eqref{eq:VFSB} this reduces to the simple condition $V_B(\fb) - V_0 =\Lambda^4$.}.
The scalar mass is then 
\begin{equation} \label{eq:FSBmass} 
    m_\phi^2 = 2 \Lambda^2 \exp\left[ -\frac{\fb^2}{\Lambda^2} \right] \, ,
\end{equation}
which is exponentially suppressed if the well size exceeds the wall thickness, $\fb\gg\Lambda$. 
For very thin walls, $\Lambda / \fb \to0$, we recover Eq.~\eqref{eq:Vwall} with a massless field. We set $V_0=-m_\phi^2\Lambda^2/2$ to ensure $V_B(0)=0$. The potential in Eq.~\eqref{eq:VFSB} is radiatively stable, yields exponentially small masses, and supports matter-like oscillations at late times. These features make $\phi$ an ideal ultralight DM candidate.

\vspace{0.1cm}

\noindent {\bf Misalignment Production.} 
The wallion equation of motion in the Friedmann–Lemaître–Robertson–Walker (FLRW) background governs its evolution. 
Initially displaced from the minimum of its potential, the field remains frozen due to Hubble friction until the restoring force becomes comparable to the Hubble damping and triggers its motion. Near the potential minimum, the oscillations can be approximated as harmonic. This constitutes the misalignment mechanism for DM production.

The homogeneous mode satisfies $\ddot{\phi}+3H\dot{\phi}+V'(\phi)=0$, where $H$ is the Hubble parameter. Overdots and primes denote FLRW time and field derivatives, respectively. We denote by $\phi_i$ the field value at time $t_i$, when the field is still held by Hubble friction.  In a radiation-dominated universe with temperature $T$, $H=\sqrt{g_\star\pi^2/90}\,T^2/\Mpl$, where $g_\star$ and $\Mpl$ are the effective number of degrees of freedom and the reduced Planck mass. The temperature $\Tosc$ at which the field begins to evolve is set by $H\simeq\sqrt{V'(\phi)/\phi}$, yielding $\Tosc\simeq\sqrt{m_\phi\Mpl}\,\exp\!\left(\phi_i^2/4\Lambda^2\right)$.

We define the dimensionless field $\ft\equiv\phi/\Lambda$ and temperature  $y\equiv\sqrt{H/m_\phi}$. For $\phi_i\lesssim\Lambda$, the field begins to move when $y\simeq1$. The equation of motion then reads~\footnote{Throughout this analysis, we approximate $g_* \simeq {\rm const}$.}
 \begin{equation} \label{eq:EOMrescaled}
    \frac{d^2 \ft}{d y^2} 
    = - \frac{1}{y^6} \frac{1}{\Lambda^2 m_\phi^2} \frac{\partial V}{\partial \ft} \,.
\end{equation}
For scalar potentials of the form $V(\phi)=m_\phi^2\Lambda^2 v(\ft)$, such as those of the wallion or axion, Eq.~\eqref{eq:EOMrescaled} is independent of $m_\phi$ and $\Lambda$. Consequently, the solution $\ft(y)$ depends only on the initial condition $\ft(y\gg1)=\phi_i/\Lambda$.
For the harmonic case, $V_{\rm HO}(\phi)=m_\phi^2\phi^2/2$, the adiabatic invariant $I_{\rm HO}\equiv\phi^2/T^3$ remains conserved during the oscillatory phase~\cite{Bae:2008ue}, which begins at $T_{\rm HO}\sim\sqrt{m_\phi\Mpl}$. Applying the virial theorem, the resulting relic abundance scales as $\Omega_{\phi}^{\rm HO}\propto\sqrt{m_\phi}\,\phi_i^2$. For a potential of the form $V(\phi)=m_\phi^2\Lambda^2 v(\ft)$ and assuming that the solution of Eq.~\eqref{eq:EOMrescaled} reaches the harmonic regime $V\simeq V_{\rm HO}$ at $y\ll1$, we can express the relic abundance of $\phi$ as
\begin{equation} \label{eq:RelicDensity_Misaligned}
      \Omega_{\phi} h^2 = \!\left( \frac{g_*}{100} \right)^{\!-1/4} \left( \frac{m_\phi}{\rm eV}\right)^{\!1/2} \left( \frac{\Lambda}{10^{12} \, {\rm GeV} }\right)^{\!2} f \!\left(\frac{\phi_i}{\Lambda} \right) .
\end{equation}
The function $f$ can be obtained from the solution $\ft(y)$ of Eq.~\eqref{eq:EOMrescaled} with initial condition $\ft(y\gg1) = \phi_i/\Lambda$, by comparing Eq.~\eqref{eq:RelicDensity_Misaligned} with $\Omega_\phi = \rho_\phi/\rho_{\rm crit} = (\dot{\phi}^2 + m_\phi^2 \phi^2)/(2 \rho_{\rm crit})$ evaluated at $T=T_0$, and it explicitly reads $f(\phi_i/\Lambda) \approx 0.16 (y^6 \ft'(y)^2 + \ft(y)^2)/y^3|_{y\ll 1}$. 
As long as $f$ is evaluated in the harmonic regime $y\ll1$~\footnote{In such a regime, we have the scalings $\ft(y)\sim y^{3/2}$ and $\ft'(y)\sim y^{-3/2}$.}, its value is independent of $y$ and determined solely by the initial $\phi_i$.
A viable wallion scenario must reproduce the observed DM relic abundance, $\Omega_{\rm DM} h^2 = 0.1200 \pm 0.0012$~\cite{Planck:2018vyg}. 

\begin{figure}
    \centering
    \includegraphics[width=0.96\linewidth]{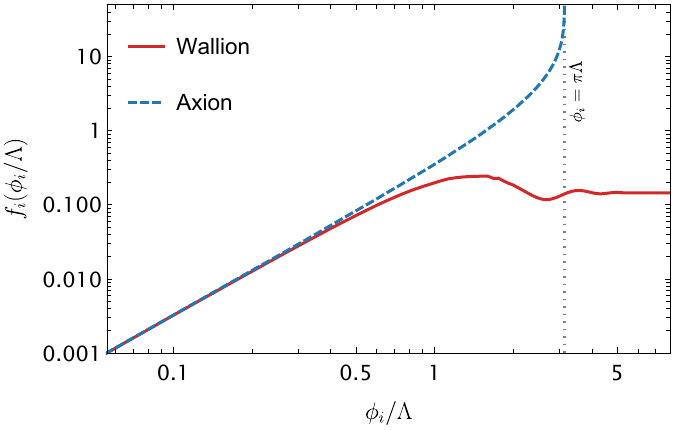}
    \caption{Numerical evaluation of $f(x)$ from Eq.~\eqref{eq:RelicDensity_Misaligned} for the  axion (dashed blue) and wallion (solid red) potentials.}
    \label{fig:FSBvsALPvsILP_geffConst}
\end{figure}

Fig.~\ref{fig:FSBvsALPvsILP_geffConst} shows numerical results for $f$ for the axion case, $V_{\rm axion}(\phi)=m_\phi^2\Lambda^2[1-\cos(\phi/\Lambda)]$, and for the potential in Eq.~\eqref{eq:VFSB}. For $\phi_i\ll\Lambda$, both are approximately harmonic, yielding $f(\phi_i/\Lambda)\simeq0.32\,(\phi_i/\Lambda)^2$. The axion displays its typical rise at $\phi_i=\pi\Lambda$. The wallion yields a smaller, eventually constant result for $\phi_i\gtrsim\Lambda$, with $f(\phi_i/\Lambda)\simeq0.14$ for $\phi_i\gtrsim4\Lambda$~\footnote{Background field values $\phi > \Lambda$ do not necessarily invalidate the EFT interpretation of $V_B(\phi)$.  The scale of new physics $\Lambda$ limits the energy density, not the field amplitude,  so the EFT remains valid as long as $V_B(\phi) < \Lambda^4$, or equivalently $\phi \lesssim \bar{\phi}$.}. At large amplitudes, the Taylor-expanded wallion potential is dominated by higher-order terms $\sim\phi^{2n}$, with $n$ increasing with $\phi_i$. In this regime, the adiabatic invariant $I_{2n}\equiv\phi^{n+1}/T^3$ is conserved, with resulting anharmonic oscillations with $\rho_\phi\propto T^{6n/(1+n)}$~\cite{Turner:1983he}. Consequently, $\sqrt{V'(\phi)/\phi}\propto T^{3(n-1)/(n+1)}$ redshifts faster than the Hubble rate for $n>5$. As such terms dominate for $\phi\gtrsim2.4\,\Lambda$, the field rolls toward $\phi\simeq2.4\,\Lambda$ before oscillating. 

\vspace{0.1cm}

\noindent {\bf On the Initial Conditions.} Cosmic inflation offers a natural origin for the initial wallion displacement. The field value at the end of inflation defines the initial condition $\phi_i$ entering Eq.~\eqref{eq:RelicDensity_Misaligned}. If the wallion potential is active during the inflationary phase, $\phi_i$ can become independent of the preinflationary value $\phi_0$, provided inflation lasts long enough to erase any memory of it. This requires a number of e-folds $N_e$ such that $\sqrt{N_e}\,(H_I/2\pi)\gg\mathrm{max}\!\left\lbrace \phi_0,\,\mathrm{min}\!\left\lbrace H_I^2/m_\phi,\,\Lambda \right\rbrace \right\rbrace$. In this \emph{stochastic} scenario, the probability distribution~\cite{Starobinsky:1994bd,Graham:2018jyp} of $\phi_i$ is given by
\begin{equation} \label{eq:distributionMis}
    P_{\rm eq}(\phi_i) \propto \exp\left[-\frac{8\pi^2}{3H_I^4}\,V(\phi_i)\right]\,. 
\end{equation}
The relic density of $\phi$ then follows from the average
\begin{equation}
    \left.\Omega_\phi h^2\right|_{\rm stochastic} 
    = \,
       \int d\phi_i \,P_{\rm eq}(\phi_i) \, \times \, \Omega_{\phi} h^2(\phi_i/\Lambda) \, ,
       \label{eq:Omegah2average}
\end{equation}
where the relic density $\Omega_\phi h^2(\phi_i/\Lambda)$ for a given initial field value $\phi_i/\Lambda$ is given in Eq.~\eqref{eq:RelicDensity_Misaligned}.

If equilibration does not occur, the resulting distribution is Gaussian with standard deviation $\sqrt{N_e}\,H_I/(2\pi)$ and mean $\phi_0$. In this \emph{locked} scenario, where the field retains memory of its inflationary value, the relic abundance of $\phi$ follows Eq.~\eqref{eq:Omegah2average} with $P_{\rm eq}(\phi_i)$ replaced by this Gaussian. If the preinflationary value is large, $\phi_0 \gg \sqrt{N_e}\,H_I/(2\pi)$, the relic density can instead be estimated by setting $\phi_i=\phi_0$ in Eq.~\eqref{eq:RelicDensity_Misaligned}.

To build intuition, we approximate the probability distribution in Eq.~\eqref{eq:distributionMis} with a window function that remains constant up to a cutoff in  field amplitude. Explicitly, $P_{\rm eq}(\phi)\propto\Theta(\phi_{\rm cut}-|\phi|)$, where $\Theta$ denotes the Heaviside step function. The cutoff is set by $V(\phi_{\rm cut})=3H_I^4/(8\pi^2)$, which for the wallion potential yields
\begin{equation}
    \frac{\phi_{\rm cut}^2}{\Lambda^2} = \ln\!\left[ 1 + \frac{3}{4 \pi^2} 
    \frac{H_I^4}{m_\phi^2 \Lambda^2} \right] \ .
\end{equation}
We identify two opposite scenarios. When $H_I \gg \sqrt{m_\phi\Lambda}$, the cutoff satisfies $\phi_{\rm cut}\gtrsim\Lambda$, and for sufficiently large $H_I$ the relic density is well approximated by Eq.~\eqref{eq:RelicDensity_Misaligned} with $f(\phi_i/\Lambda\!\gg\!1)\simeq0.14$. Requiring $\Lambda$ values consistent with the observed relic abundance, this regime occurs for $H_I \gg 30 \,(m_\phi/{\rm eV})^{3/8}\,{\rm GeV}$. Conversely, the cutoff satisfies $\phi_{\rm cut}\ll\Lambda$ when $H_I\ll\sqrt{m_\phi\Lambda}$. In this regime, the harmonic approximation holds, giving $f(\phi_i/\Lambda)\simeq0.32\,(\phi_i/\Lambda)^2$. For $g_\star=100$, the resulting relic abundance is $\Omega_\phi h^2\simeq 8.1 \times10^{-9}\,(H_I/{\rm GeV})^4\,(m_\phi/{\rm eV})^{-3/2}$. This corresponds to $m_\phi\simeq 1.7 \times 10^{-5}\,(H_I/{\rm GeV})^{8/3}\,{\rm eV}$, reproducing the well-known result for a free massive scalar~\cite{Tenkanen:2019aij}.

\vspace{0.1cm}
\noindent
\textbf{Isocurvature.} The wallion can generate isocurvature fluctuations since it is decoupled from radiation and unsourced by inflaton perturbations. Cosmic Microwave Background (CMB) data constrain their power to $\mathcal{P}_S < 7.6\times10^{-11}$ at the pivot scale $k=0.002\,{\rm Mpc}^{-1}$~\cite{Planck:2018vyg}.

In the locked scenario, the isocurvature power spectrum can be approximated as $\mathcal{P}_S \approx \langle (\rho_\phi - \langle \rho_\phi \rangle)^2/\langle \rho_\phi \rangle^2 \rangle$, where the average is taken over a Gaussian distribution with mean $\phi_i$ and standard deviation $H_I/(2\pi)$~\cite{Hamann:2009yf,Hertzberg:2008wr,Visinelli:2009zm,Graham:2025iwx}. 
For initial misalignments in the range $H_I/(2\pi) \ll \phi_i \lesssim \Lambda$, the CMB constraint implies an upper bound on the inflationary scale, $H_I/{\rm GeV}\lesssim10^7\,(m_\phi/{\rm eV})^{-1/4}$, consistent with~\cite{Tenkanen:2019aij}. 
The opposite limit, $\phi_i<H_I/(2\pi)$, is excluded for $\phi_i\lesssim\Lambda$ since $\mathcal{P}_S\simeq2$. Conversely, large misalignments $\phi_i \gtrsim \Lambda$ always evade the isocurvature bound in the locked scenario, as $f(\phi_i/\Lambda)$ saturates to a constant.

For the stochastic scenario, the isocurvature power spectrum is computed following the approach of~\cite{Starobinsky:1994bd,Markkanen:2019kpv}. When $H_I^2\ll\Lambda m_\phi$, existing results for quadratic potentials apply~\cite{Tenkanen:2019aij}, whereas for $H_I^2\gtrsim\Lambda m_\phi$ we numerically solve the Fokker--Planck equation for the wallion potential~\footnote{The spectrum is obtained as a sum over modes with increasing spectral index. Unlike the quadratic case, higher-order terms contribute, and we observe numerical convergence after $\sim25$ modes.}. We find that $m_\phi\gtrsim0.5\,{\rm GeV}$ is excluded by isocurvature constraints unless $H_I\gtrsim10^{10}\,{\rm GeV}$.

\begin{figure}
    \centering
    \includegraphics[width=0.96\linewidth]{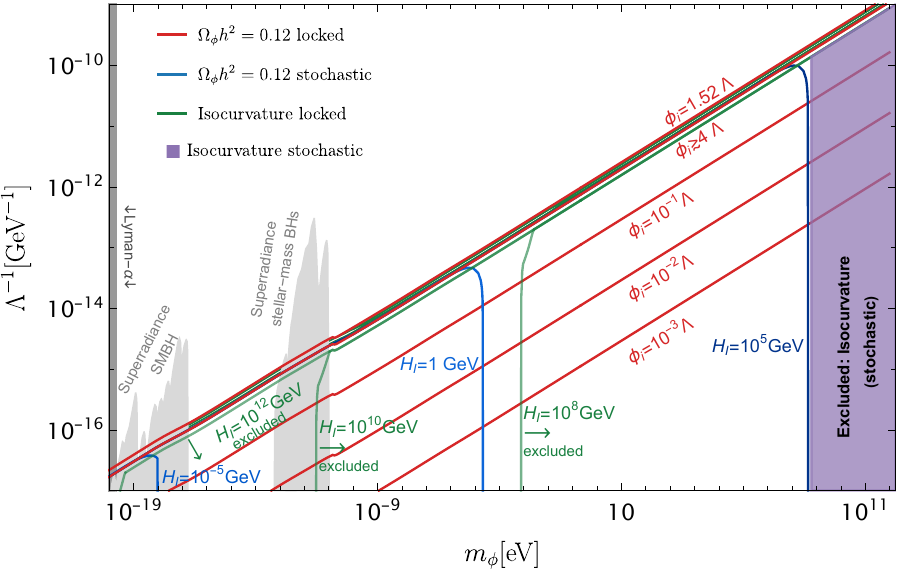
    }
    \caption{Wallion parameter space in the $(m_\phi,\Lambda^{-1})$ plane. See text for a detailed description.}
    \label{fig:FSBduringInflationPred}
\end{figure}

\vspace{0.1cm}
\noindent
\textbf{Scenarios for Wallion Cosmology.} 
In the $(m_\phi,\Lambda^{-1})$ plane, solid red and blue lines show $\Omega_\phi = \Omega_{\rm DM}$ isocontours for the locked and stochastic scenarios, respectively~\footnote{In Fig.~\ref{fig:FSBduringInflationPred}, we have $6 \lesssim \fb / \Lambda \lesssim 15$ (see Eq.~\eqref{eq:FSBmass}).}. In Fig.~\ref{fig:FSBduringInflationPred}, we summarize the cosmologically viable parameter space in the $(m_\phi,\Lambda^{-1})$ plane. Solid red and blue lines show $\Omega_\phi = \Omega_{\rm DM}$ isocontours for the locked and stochastic scenarios, respectively. The blue lines converge to the red line for $\phi_i \gtrsim 4 \Lambda$ for $H_I\!\gtrsim\!10^3\,\mathrm{GeV}\left( m_\phi/\mathrm{eV}\right)^{3/8}$. 
The field space boundary $\fb$, related to $(m_\phi,\Lambda^{-1})$ through Eq.~\eqref{eq:FSBmass}, takes values in the range $6 \lesssim \fb / \Lambda \lesssim 15$. Since $\Omega_\phi$ is independent of $\phi_i$ if $\phi_i \gtrsim 4 \Lambda$, all points in Fig.~\ref{fig:FSBduringInflationPred} can be realized satisfying $\phi_i < \fb$. Similarly, as long as $H_I \lesssim \Lambda$, the stochastic scenario satisfies $\phi_\text{cut} \lesssim \fb$.
We also identify regions excluded by isocurvature constraints. For the locked scenario, we show bounds for different values of $H_I$ with the region on the right of the colored green lines being excluded. For the stochastic scenario, the value of $H_I$ for each point in the plane is identifed by the relic density constraint, and this leads to the exclusion of the purple region~\footnote{To ease the visualization, we interrupt the isocurvature exclusion lines before they approach the red one for $\phi_i \gtrsim 4 \Lambda$ which is not subject to isocurvature bounds.}. Shaded gray regions show constraints from the Lyman-$\alpha$ forest~\cite{Kobayashi:2017jcf,Rogers:2020ltq} and black-hole superradiance~\cite{Stott:2018opm,Davoudiasl:2019nlo,Hoof:2024quk,Witte:2024drg,Caputo:2025oap,Unal:2020jiy}. The latter are taken from Refs.~\cite{Witte:2024drg,Caputo:2025oap,Unal:2020jiy}. LISA could probe $m_\phi \in \left[10^{-13},10^{-15}\right] \,\mathrm{eV}$~\cite{Kim:2025wwj}.
Superradiance bounds cease above a certain $\Lambda^{-1}$\cite{Baryakhtar:2020gao} due to efficient $\phi$ self-interactions~\footnote{The quartic wallion self-interaction is repulsive, unlike for axions, which may be relevant for repulsive/superfluid DM~\cite{Guth:2014hsa,Ferreira:2020fam,Berezhiani:2025maf}.}.
Ultrafaint dwarf–galaxy modeling (not shown) provides further limits, the strongest implying $m_\phi \gtrsim 8\times10^{-18}\,\mathrm{eV}$~\cite{Dalal:2022rmp,May:2025ppj}.

\vspace{0.1cm}

\noindent
\textbf{A Plausible Origin: Instantons.} We now turn to the microscopic origin of the wallion potential, showing that it can arise from instanton dynamics. Our approach parallels Ref.~\cite{Davoudiasl:2024epi}, where the Higgs potential was generated through instanton effects. We extend the SM by introducing a dark gauge group $SU(N)$ and a scalar field $\phi$. The Lagrangian includes the following two operators:
\begin{equation}
    \mathcal{L} \supset  -\frac{1}{4 g_d^2} G_{\mu \nu} G^{\mu \nu} + \frac{1}{4 M^2} \phi^2 G_{\mu \nu} G^{\mu \nu} \,,
    \label{eq:Linst}
\end{equation}
where $g_d$ and $G_{\mu \nu}$ denote the dark gauge coupling and field-strength. The first term is the kinetic operator for the dark gauge bosons, while the dimension-6 contact interaction effectively redefines the gauge coupling
\begin{equation} \label{eq:geff}
    \frac{1}{g_{\rm eff}^2(\mu,\phi)} = \frac{1}{g_d^2(\mu)} - \frac{\phi^2}{M^2} \,,
\end{equation}
where $\mu$ is the renormalization scale. Nonperturbative instanton effects can then generate the potential~\cite{Belavin:1975fg,tHooft:1976rip,Jackiw:1976pf,Callan:1976je,Marino:2015yie}
\begin{equation} \label{eq:Vinstanton}
    V_{\rm inst}(\phi) = K(\mu) \exp\!\left[-\frac{8\pi^2}{g_{\rm eff}^2(\mu, \phi)} \right] \,.
\end{equation}
Up to the constant $V_0$, this reproduces the wallion potential in Eq.~\eqref{eq:VFSB} with $\fb^2 = M^2/g^2(\mu) - \ln\!\left[K(\mu)/\Lambda^4\right]\Lambda^2$ and $\Lambda^2 = M^2/(8\pi^2)$. For $M^2>0$, the Lagrangian in Eq.~\eqref{eq:Linst} thus induces a field-space boundary at $|\phi|=\fb$, beyond which the effective gauge coupling squared $g_{\rm eff}^2$ defined in Eq.~\eqref{eq:geff} becomes negative. The sign and magnitude of $K(\mu)$ are determined by the instanton dynamics of the dark $SU(N)$. The existence of the boundary and the stability of the potential both require $K(\mu)>0$.

We briefly comment on possible UV completions that generate the higher-dimensional operator in Eq.~\eqref{eq:Linst}. One possibility is a mild breaking of the shift symmetry through interactions such as $\lambda\phi^2\eta^\dagger\eta$ or $\lambda\phi\bar{\psi}_1\psi_2$, where $\eta$ and $\psi_i$ are $SU(N)$-charged scalars and fermions, respectively. In the fermionic case, the linear coupling in $\phi$ can be forbidden by assigning suitable charges to $\phi$ and $\psi_i$ under, for instance, a $\mathbb{Z}_2$ symmetry.


\vspace{0.1cm}
\noindent
\textbf{On the Possibility of SM interactions.} All observable effects discussed so far, including the relic density and isocurvature signals, arise solely from gravitational interactions and wallion self-interactions encoded in the scalar potential of Eq.~\eqref{eq:VFSB}. Although couplings to SM fields are not required for the internal consistency of the framework, they open the appealing possibility of wallion detection in terrestrial experiments. If the heavy degrees of freedom generating the dimension-6 operator in Eq.~\eqref{eq:Linst} carry SM charges, additional higher-dimensional operators naturally emerge, such as
\begin{equation} \label{eq:ULDMFFcoupling}
    \mathcal{L} \supset \frac{c_F}{4 M^2} \phi^2 F_{\mu \nu} F^{\mu \nu} \, ,
\end{equation}
where $c_F$ is a coefficient determined by the UV completion and $F_{\mu\nu}$ denotes the electromagnetic field strength. This operator calls for a reassessment of the relic density analysis and threatens the naturalness of the wallion mass. In what follows, we focus on this example, keeping in mind that couplings to other SM fields may also arise. 

Operators such as Eqs.~\eqref{eq:Linst} and~\eqref{eq:ULDMFFcoupling} induce radiative corrections to the wallion mass. At one loop, this contribution is estimated as $m_{\phi,\,\text{1-loop}}^2 \sim \Lambda_\text{UV}^4 / (512 \pi^4 \Lambda^2)$\footnote{This estimate takes the coefficient of the operator in Eq.~\eqref{eq:ULDMFFcoupling} with a loop suppression, uses the identification between $M$ and $\Lambda$, and fixes the UV cutoff dependence by dimensional analysis.}, where $\Lambda_\text{UV}$ denotes a UV cutoff. Such corrections are generic in models with quadratically coupled ultralight DM~\cite{Banerjee:2022sqg,Delaunay:2025pho,Brzeminski:2020uhm}. We regard the instanton-induced potential as natural if $m_\phi^2 > m_{\phi,\,\text{1-loop}}^2$. This is easily met when $\phi$ is decoupled from the SM, since $\Lambda_\text{UV}$ can be arbitrarily low. In contrast, once Eq.~\eqref{eq:ULDMFFcoupling} is included, collider limits require $\Lambda_\text{UV}\gtrsim100~\mathrm{GeV}$. Whether this correction actually arises depends on the UV completion; for example, in frameworks such as Ref.~\cite{Delaunay:2025pho}, inspired by twin-Higgs models~\cite{Chacko:2005pe}, the first nonvanishing contribution appears only at two loops. Thus, the naturalness issue is UV-sensitive rather than intrinsic to the wallion framework.

Interactions with photons induce a thermal contribution to the wallion mass~\cite{Bouley:2022eer}
\begin{equation} \label{eq:thermalmass}
    \Delta m^2_{\phi}(T) = c_F \alpha_\text{em}^2 \frac{\pi^2}{3} \frac{T^4}{M^2} \, ,
\end{equation}
where $\alpha_\text{em}$ is the fine-structure constant~\footnote{Ref.~\cite{Bouley:2022eer} derives the thermal mass for an electron plasma, while in general all charged species can contribute~\cite{Cyncynates:2024bxw,Cyncynates:2024ufu,Baryakhtar:2025uxs}. A future study of the thermal history should include these effects but we do not expect them to qualitatively alter our conclusions.}. In a UV completion with a quartic coupling between the wallion and a heavy charged scalar, $\lambda \phi^2 \eta^\dagger \eta$, the dimension-6 operators in Eqs.~\eqref{eq:Linst} and~\eqref{eq:ULDMFFcoupling} acquire the same sign. Because the presence of a field-space boundary requires this sign to be opposite to that of the gauge kinetic term, the resulting squared thermal mass of the wallion is positive.

We assume instantaneous reheating with reheat temperature $\Trh$ and set $H_I = \sqrt{g_\star\pi^2/90}\,\Trh^2/\Mpl$. We further consider inflaton decays exclusively into SM states and restrict to $\Trh < M$. Consequently, the dark sector is cold at reheating, and the wallion potential is present throughout. Moreover, dark gluons never thermalize and do not contribute to dark radiation.

The wallion evolution is governed by two different contributions: the potential with boundaries in Eq.~\eqref{eq:VFSB} and the thermal mass in Eq.~\eqref{eq:thermalmass}. We first consider the locked inflationary scenario, with $\phi(\Trh)=\phi_i$ and the thermal mass dominating at $\Trh$. Assuming also $\Delta m_\phi(T_{\rm rh}) > H(T_{\rm rh})$, the field undergoes harmonic oscillations with a temperature-dependent mass, as the QCD axion before confinement but following the scaling in Eq.~\eqref{eq:thermalmass}. The ratio $\Delta m_\phi(T)/H(T)$ is temperature independent, so this regime occurs only if $\Delta m_\phi(T_{\rm rh}) > H(T_{\rm rh})$, corresponding to $\Lambda^{-1}\gtrsim 8.7 \times10^{-17}\sqrt{g_\star/c_F}\,\mathrm{GeV}^{-1}$. The invariant $I_T\equiv m_\phi(T)\phi(T)^2T^{-3}$ implies the subsequent redshift $\phi\propto T^{1/2}$. Eventually, the vacuum potential becomes dominant. Provided the preceding oscillatory phase persists long enough, the transition occurs when the wallion potential is effectively harmonic. Accordingly, the transition occurs at the temperature $T_c$ defined by $\Delta m_\phi(T_c)=m_\phi$, with $m_\phi$ from Eq.~\eqref{eq:FSBmass}. This is consistent provided $\phi_i/\Lambda \ll (\Trh/T_c)^{1/2}$. The subsequent evolution is matterlike, with a field amplitude at matter–radiation equality estimated as $\phi(T_{\rm eq}) \simeq (T_{\rm eq}/T_c)^{3/2}(T_c/\Trh)^{1/2}\phi_i$. Matching this to the radiation energy density at that epoch, $m_\phi^2\phi(T_{\rm eq})^2 = g_\star(T_{\rm eq})(\pi^2/30)\Teq^4$, shows that the wallion reproduces the observed DM relic abundance for $\Lambda^{-1} \simeq 2\times10^{-3}\sqrt{c_F}(m_\phi/{\rm eV})(\phi_i/\Lambda)^2\Trh^{-1}$.

There are exceptions to the situation described above. First, the thermal mass may be too small to permit oscillations ($\Delta m_\phi(\Trh) < H(\Trh)$), with the field frozen until the wallion potential triggers its motion and the relic abundance given by Eq.~\eqref{eq:RelicDensity_Misaligned}, provided $\Trh>T_\star$. Futhermore, $\Trh$ can be small enough such that the wallion potential always dominates over the thermal mass. Assuming the field at reheating satisfies $\phi_i \lesssim \Lambda$, the harmonic part of the potential drives its motion. If we have $\Trh > T_\star$, the DM relic density constraint reads $\Lambda^{-1} \simeq (m_\phi/{\rm eV}) (\phi_i/\Lambda) (T_{\rm eq}/T_{\rm rh})^{1/2} T_{\rm rh}^{-1}$. Finally, we address the stochastic scenario. Crucially, it is not possible to reproduce the relic density when $\Delta m_\phi(\Trh) > H(\Trh)$ since the existence of the wallion potential during inflation requires the Hubble parameter $H_I$ to be smaller than the dark confinement scale $\Lambda_D \simeq \sqrt{m_\phi \Lambda}$. In the opposite case, the results for the stochastic scenario presented in Fig.~\ref{fig:FSBduringInflationPred} apply, as long as $\Trh > T_\star$ and $H_I \lesssim \Lambda_D$.

\begin{figure}
    \centering
    \includegraphics[width=0.96\linewidth]{
    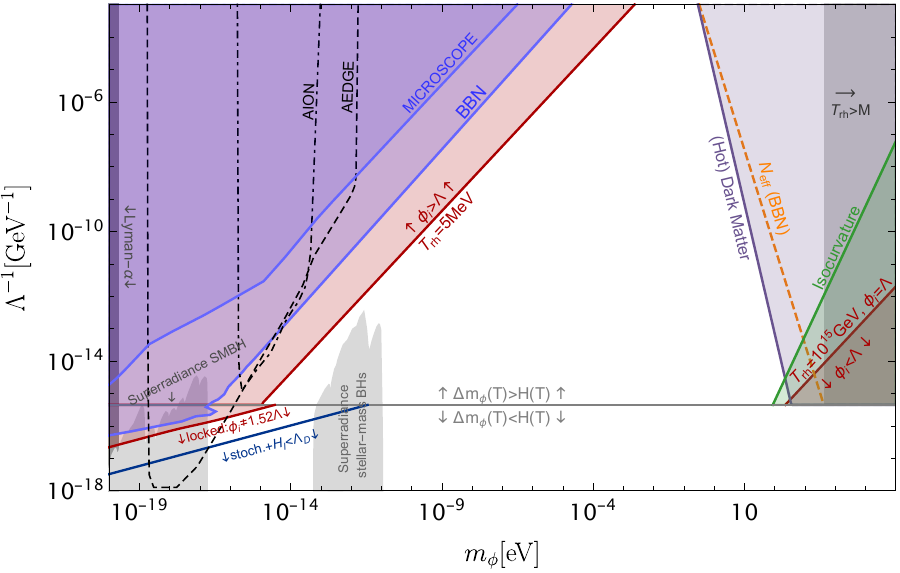
    }
    \caption{Phenomenology of wallion-photon interactions in the $(m_\phi,\Lambda^{-1})$ plane. See text for a detailed description.  
    }
    \label{fig:ExpLimitsUVModel}
\end{figure}

Fig.~\ref{fig:ExpLimitsUVModel} summarizes the phenomenological constraints in the $(m_\phi, \Lambda^{-1})$ plane, assuming $c_F=1$. Above the horizontal gray line, identified by $\Delta m_\phi(T_{\rm rh}) = H(T_{\rm rh})$, we consider DM production driven by $\Delta m_\phi (T)$ in the locked scenario and $\Trh > T_c$. We set $\phi_i=\Lambda$ and at each point determine $\Trh$ reproducing $\Omega_\phi = \Omega_{\rm DM}$. The solid red lines identify $\Trh=5 \, \mathrm{MeV}$ and $\Trh=10^{15} \, \mathrm{GeV}$. The region to the left of the $\Trh=5 \, \mathrm{MeV}$ line requires larger initial misalignment $\phi_i$ because of the Big Bang Nucleosynthesis (BBN) constraints~\cite{Hannestad:2004px,deSalas:2015glj,Barbieri:2025moq}. Contrarily, the region to the right of the $\Trh=10^{15} \, \mathrm{GeV}$ line requires smaller  $\phi_i$ due to constraints on the energy scale of inflation~\cite{Planck:2018vyg}. The green shaded region indicates exclusion from isocurvature once we impose on each point a value of $\Trh$ that reproduces the relic density for $\phi_i=\Lambda$. 

Thermal scattering of bath particles produces wallions that can contribute to dark radiation or a hot DM subcomponent. Their production rate per unit volume is $\gamma_\phi(T) = (3 c_F^2 / \pi^3) T^8 / M^4$, yielding an asymptotic comoving number density $Y_\phi^\infty \simeq 0.2 \, (c_F^2 / g_\star^{3/2}) (M_{\rm Pl} \Trh^3 / M^4)$. Dark radiation is constrained by BBN~\cite{Yeh:2020mgl,Pisanti:2020efz,Yeh:2022heq} (for $m_\phi \lesssim 1\,{\rm MeV}$) and by the CMB~\cite{Planck:2018vyg} (for $m_\phi \lesssim 0.1\,{\rm eV}$), both requiring $\Delta N_{\rm eff} \lesssim 0.3$. Following Ref.~\cite{DEramo:2023nzt}, we obtain $\Delta N_{\rm eff} \simeq 75.6 (Y_\phi^\infty)^{4/3}$\footnote{This procedure suffices for current constraints, while future CMB surveys will require a dedicated phase-space analysis (see, e.g., Refs.~\cite{DEramo:2024jhn,Badziak:2024qjg}).}, which in turn implies
$\Lambda^{-1} \lesssim 4.6 \, (g_\star^{3/8} / c_F^{1/2}) M_{\rm Pl}^{-1/4} \Trh^{-3/4}$. For $m_\phi \gtrsim 0.1 \, {\rm eV}$, thermally produced wallions form a subdominant hot DM component. Requiring their mass density to stay below one percent of the observed DM abundance~\cite{DEramo:2020gpr} yields
$\Lambda^{-1} \lesssim 3.3 \, (g_\star^{3/8} / c_F^{1/2}) ({\rm eV}/m_\phi)^{1/4} M_{\rm Pl}^{-1/4} \Trh^{-3/4}$.

Ultralight DM featuring a quadratic coupling to photons can be explored through a broad suite of experimental searches~\cite{Antypas:2022asj,Banerjee:2022sqg,Bouley:2022eer,Beadle:2023flm,Kim:2023pvt}. We show in Fig.~\ref{fig:ExpLimitsUVModel} with blue shaded regions constraints from tests of the weak equivalence principle~\cite{Schlamminger:2007ht,MICROSCOPE:2019jix}, time-variation in atomic transition frequencies~\cite{Arvanitaki:2014faa}, BBN~\cite{Sibiryakov:2020eir,Bouley:2022eer,Ghosh:2025pbn}. We also show with black dashed lines projections from upcoming atom interferometers~\cite{AEDGE:2019nxb,Badurina:2019hst}. While not presented here, recent limits from sourcing in white dwarfs may also apply~\cite{Bartnick:2025lbg}. 

Finally, we discuss the region below the horizontal gray line where thermal effects are negligible. The red and blue curves denote the regions yielding successful DM production in the locked and stochastic inflationary scenarios, respectively. 
Each line corresponds to the maximal achievable coupling $\Lambda^{-1}$ in the given scenario, obtained for $\phi_i \simeq 1.52\Lambda$ (locked) and $H_I = \Lambda_D$ (stochastic).

\vspace{0.1cm}

\noindent {\bf Conclusions.} In this work, we introduced the wallion field with a potential bounded in field space and an exponentially suppressed and radiatively stable mass. The same potential drives matter-like oscillations at late times, making the wallion a compelling ultralight DM candidate. Unlike axion models with periodic potentials, the present-day energy density becomes independent of the inflationary misalignment once this exceeds a critical value. Intriguingly, this property is responsible for the suppression of isocurvature perturbations.

We explored the possible origin from instanton effects in a confining dark gauge sector. This requires a dimension-6 quadratic coupling between the wallion and the gauge kinetic term. Remarkably, the framework operates entirely within the dark sector, without invoking interactions with SM fields. The framework becomes even more intriguing if the wallion also couples to visible particles. Such interactions can naturally emerge within the instanton setup when the heavy states responsible for the dimension-6 operator carry SM gauge charges. As we have shown, these couplings influence the cosmological evolution and relic density, while simultaneously opening a window for detection in forthcoming precision experiments such as AION and AEDGE.

This work sets the stage for several future developments. From the model-building perspective, a systematic study of UV completions that realize instanton-generated field-space boundaries represents an important next step. The inclusion of SM interactions enriches the early-universe dynamics, and a detailed treatment will be essential to fully establish the framework. In a complementary direction, one may explore alternative origins of field-space boundaries, for instance from non-canonical kinetic terms in extra-dimensional constructions or from potentials induced for pseudo–Nambu–Goldstone bosons of non-compact, softly broken symmetries. These ideas are not limited to the exponential potential considered here but may apply more broadly to any setup that effectively behaves as an infinite potential well. From a phenomenological standpoint, field-space boundaries may also have far-reaching implications, including roles in inflationary model building, distinctive signatures in primordial non-Gaussianities, or a possible link to superfluid DM via the induced repulsive quartic self-interaction. Altogether, these features identify field-space boundaries as a new and testable guiding principle for the origin of DM and scalar field dynamics in the early universe.

\vspace{0.1cm}
\noindent {\it Acknowledgements.} The authors thank Clifford Cheung, Jun'ya Kume, Alessandro Lenoci, Martin Mojahed, Hitoshi Murayama, Clara Murgui, Pablo Olgoso, Sebastian Schenk, Pedro Schwaller, Nudžeim Selimović, Samuel Witte, Lorenzo Ubaldi, and Tien-Tien Yu for valuable feedback. This work was supported by Istituto Nazionale di Fisica Nucleare (INFN) through the Theoretical Astroparticle Physics (TAsP) project, and in part by the Italian MUR Departments of Excellence grant 2023-2027 “Quantum Frontiers”. M.B. was supported in part by the Italian Ministry of University and Research (MUR) through the PRIN 2022 project n. 20228WHTYC (CUP: I53C24002320006 and C53C24000760006). V.V.~was supported by the European Union's Horizon Europe research and innovation program under the Marie Sk\l{}odowska-Curie grant agreement No.~101065736, the Estonian Research Council grant RVTT7 and the Center of Excellence program TK202.

\bibliographystyle{apsrev4-2}
\bibliography{biblio.bib}

\end{document}